\documentclass[pra,superscriptaddress,amsmath,amssymb]{revtex4}

\usepackage{graphicx}
\usepackage{subfigure}
\usepackage{amsmath}
\usepackage{amssymb}
\usepackage{amsfonts}
\usepackage{euscript}
\usepackage{pslatex}
\usepackage{color}
\usepackage{hyperref}

\newcommand{\Qopi}{Q_{\mathrm{op,i}}}

\newcommand{\Qmi}{Q_{\mathrm{m,i}}}

\newcommand{\meff}{m_{\mathrm{eff}}}
\newcommand{\Vm}{V_\mathrm{m}}
\newcommand{\Vo}{V_\mathrm{o}}
\newcommand{\Leff}{L_{\mathrm{OM}}}

\newcommand{\num}{\nu_{\mathrm{m}}}
\newcommand{\nuo}{\nu_{\mathrm{o}}}
\newcommand{\Omegam}{\Omega_{\mathrm{m}}}
\newcommand{\omegao}{\omega_{\mathrm{o}}}

\newcommand{\kappaext}{\kappa_{\mathrm{ext}}}
\newcommand{\kappai}{\kappa_{\mathrm{i}}}

\newcommand{\Vac}{V_{\mathrm{acoustic}}}

\newcommand{\ddt}{\frac{\mathrm{d}}{\mathrm{d}t}}

\newcommand{\modulus}[1]{\left|#1\right|}
\newcommand{\Imag}[1]{\mathrm{Im}\left\{#1\right\}}
\renewcommand{\Re}[1]{\mathrm{Re}\left\{#1\right\}}
\newcommand{\deriv}[2]{\frac{\mathrm{d}#1}{\mathrm{d}#2}}

\newcommand{\bv}[1]{\mathbf{#1}}
\newcommand{\maxvec}[1]{\textrm{max}\!\left( \left| \mathbf{#1} \right| \right)}

\begin{document}

\pagenumbering{arabic}

\title{Optomechanical Crystals}

\author{Matt Eichenfield}
\author{Jasper Chan}
\author{Ryan M. Camacho}
\author{Kerry J. Vahala}
\author{Oskar Painter}
\affiliation{Thomas J. Watson, Sr, Laboratory of Applied Physics, California Institute of Technology, Pasadena, California 91125, USA}
\email[e-mail:  ]{opainter@caltech.edu}

\begin{abstract}

\noindent Structured, periodic optical materials can be used to form photonic crystals capable of dispersing, routing, and trapping light.  A similar phenomena in periodic elastic structures can be used to manipulate mechanical vibrations.  Here we present the design and experimental realization of strongly coupled optical and mechanical modes in a planar, periodic nanostructure on a silicon chip.  200-Terahertz photons are co-localized with mechanical modes of Gigahertz frequency and 100-femtogram mass.  The effective coupling length, $\Leff$, which describes the strength of the photon-phonon interaction, is as small as 2.9~$\mu$m, which, together with minute oscillator mass, allows all-optical actuation and transduction of nanomechanical motion with near quantum-limited sensitivity.  Optomechanical crystals have many potential applications, from RF-over-optical communication to the study of quantum effects in mesoscopic mechanical systems.  
\end{abstract}

\maketitle

Periodicity in materials yields interesting and useful phenomena.  Applied to the propagation of light, periodicity gives rise to photonic crystals\cite{ref:Yablonovitch_original_PGB}, which can be precisely engineered to, among other things, transport and control the dispersion of light\cite{ref:PCF_bandgap,ref:Notomi_large_gv_dispersion_pc_waveguide}, tightly confine and trap light resonantly\cite{ref:Noda_high_Q_PC}, and enhance nonlinear optical interactions\cite{ref:Soljacic02}.  Photonic crystals can also be formed into planar lightwave circuits for the integration of optical and electrical microsystems\cite{ref:McNab03}.  Periodicity applied to mechanical vibrations yields phononic crystals, which harness mechanical vibrations in a similar manner to optical waves in photonic crystals\cite{ref:review_phononic_crystal_apps,ref:Kushwaha_acoustic_bands,ref:Espinosa_ultrasonic_band_gap,ref:sound_attenuation_2d_array,ref:Robertson_acoustic_stop_bands,ref:phononic_crystal_waveguide,ref:psc_circuits_theory}.  As has been demonstrated in studies of Raman scattering in epitaxially grown vertical cavity structures\cite{ref:Trigo02} and photonic crystal fibers\cite{ref:pcf_om}, the simultaneous confinement of mechanical and optical modes in periodic structures can lead to greatly enhanced light-matter interactions.  A logical next step is thus to create planar circuits that act as both photonic \emph{and} phononic crystals\cite{ref:Maldovan_Simultaneous_Localization}:  optomechanical crystals.  In this spirit, we describe the design, fabrication, and characterization of a planar, silicon-chip-based optomechanical crystal capable of co-localizing and strongly coupling 200 Terahertz photons and 2 Gigahertz phonons.  These planar optomechanical crystals bring the powerful techniques of optics and photonic crystals to bear on phononic crystals, providing exquisitely sensitive (near quantum-limited), optical measurements of mechanical vibrations, while simultaneously providing strong non-linear interactions for optics in a large and technologically-relevant range of frequencies.


\begin{figure}[b]
\begin{center}
\includegraphics[width=0.6\columnwidth]{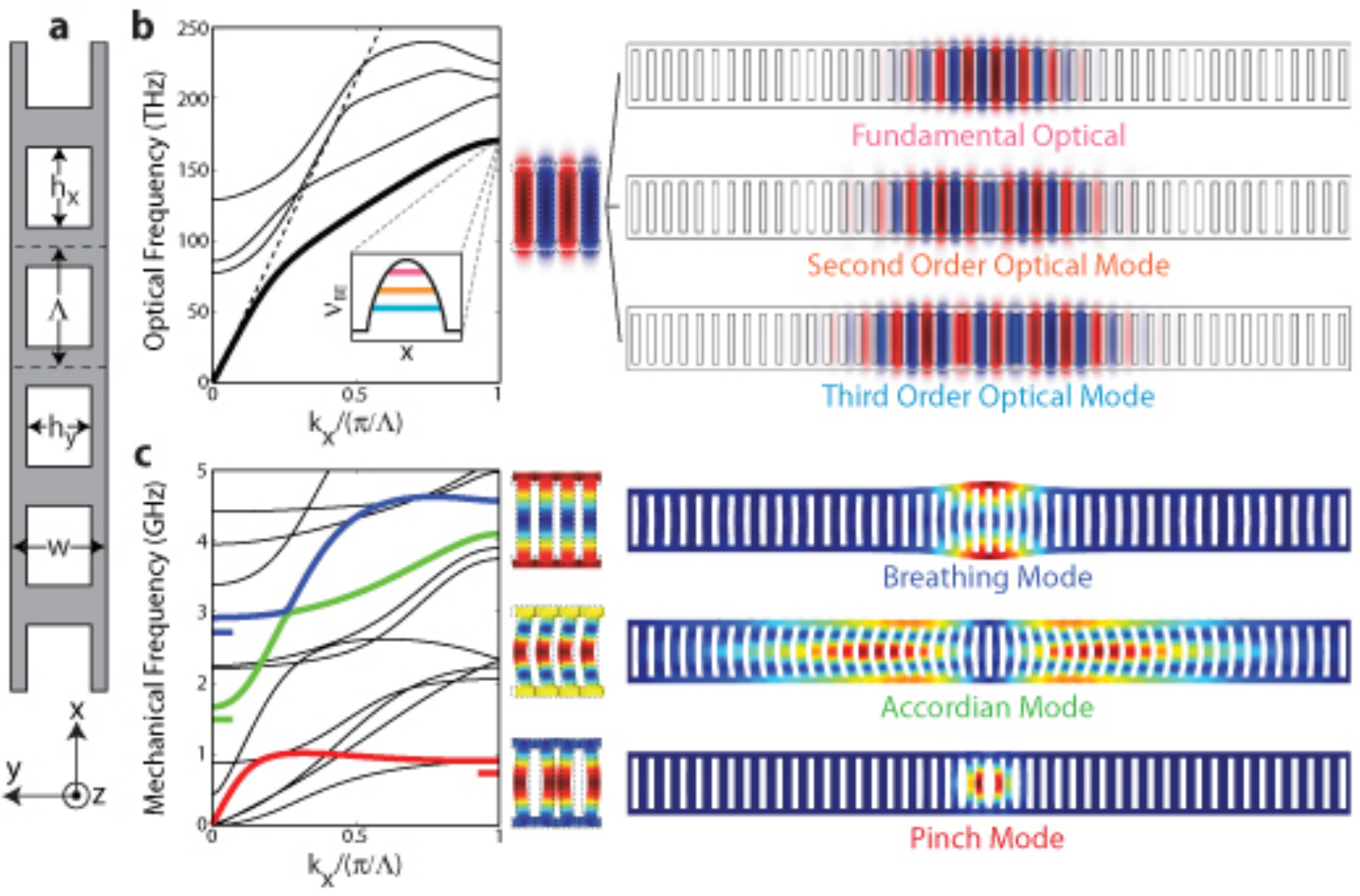}
\caption{\textbf{Optomechanical crystal design}.  \textbf{a}, Geometry of nanobeam structure. \textbf{b}, Optical and \textbf{c}, mechanical bands and defect modes calculated via FEM for the projection of the experimentally-fabricated silicon nanobeam ($\Lambda = 362$ nm, $w = 1396$ nm, $h_y = 992$ nm, $h_x = 190$ nm, and $t = 220$ nm; isotropic Young's modulus of $168.5$ GPa; $n = $3.493).  In this particular structure, which will be referred to as ``Device 1", $N_{\mathrm{defect}}= 15$ holes, $N_{\mathrm{total}}= 75$ and the spacing between the holes varies quadratically from the lattice constant of the projection (362 nm) to 85\% of that value (a ``15\% defect") for the two holes straddling the central cross-bar (the other parameters of Device 1 are as listed above).} \label{fig:bands}
\end{center}
\end{figure}

The geometry of the optomechanical crystal structure considered here is shown in Fig.~\ref{fig:bands}a.  The effectively one-dimensional (1D) optomechanical crystal consists of a silicon nanobeam (thickness $t$ not shown) with rectangular holes and thin cross-bars connected on both sides by thin rails (we will refer to infinitely periodic constructs such as this as the ``projection" of the finite structure).  Fig.~\ref{fig:bands}b shows a finite-element-method (FEM) simulation of the optical band structure of the projection of a nanobeam (see caption for parameters).  The electric field profile for modes at the  band edge ($k_x = \pi / \Lambda$, the boundary of first Brillouin zone) are shown to the right of the band structure.  The finite structure terminates at its supports on both ends, forming a doubly-clamped beam.  To form localized resonances in the center of the structure, the discrete translational symmetry of the patterned beam is intentionally disrupted by a ``defect", consisting of a quadratic decrease in the lattice constant, $\Lambda$, symmetric about the center of the beam for some odd number of holes, $N_{\mathrm{defect}}<N_{\mathrm{total}}$.  The defect forms an effective potential for optical modes at the band edges, with the spatial dependence of the effective potential closely following the spatial properties of the defect\cite{ref:ChanJ1} (as illustrated in the inset of the optical band diagram).  Thus the optical modes of the infinitely-periodic structure are confined by a quasi-harmonic potential.  This effective potential localizes a ``ladder" of modes with Hermite-Gauss envelopes, analogous to the modes of the 1D harmonic potential of quantum mechanics.  The localized optical modes of the finite structure (hereafter referred to as Device 1) are also found by FEM simulation and shown in Fig.~\ref{fig:bands}b to the right of the corresponding mode of the projection.

Analogously, Fig.~\ref{fig:bands}c shows a FEM simulation of the mechanical band structure of the nanobeam's projection.  Mechanical modes at the band edge experience an effective potential analogous to the optical modes, localizing certain types of vibrations to the defect region.  The colored bands give rise to mechanical modes that, when localized by the defect, yield ``ladders" of modes with strong dispersive coupling to the localized optical modes (the frequency of the fundamental defect mode is indicated by a horizontal bar of the same color).  The red mechanical band is analogous to the acoustic (longitudinal) vibrations of a solid or system of masses and springs, starting off at $k_x=0$ as a pure translation of the entire structure and evolving to differential acoustic vibrations of every pair of cross-beams at the band edge (mode at $k_x = \pi / \Lambda$ shown to right of band diagram), just as the highest-frequency acoustic vibrations of a solid have every unit cell vibrating out of phase with its nearest neighbors.  The green band also involves in-plane vibrations of the cross beams, but, in this case, the rails recoil in opposition to the cross-beam vibration, which means that the structure can vibrate (rather than translate) even at very long-wavelengths, giving rise to a non-zero frequency at $k_x=0$.  The blue band consists of in-plane vibrations transverse to the axis of the structure, where, at $\Gamma$ (mode at $\Gamma$ shown to right) the structure appears to ``breathe", as the cross-beams are stretched and compressed.  We classify these optomechanically-coupled mechanical modes, from lowest to highest frequency, as ``pinch'', ``accordian'', and ``breathing'' modes.  The localized mechanical modes of Device 1 are shown to the right of the corresponding mode of the projection.  

The two kinds of waves, mechanical and optical, are on equal footing in this structure.  Each mechanical mode has a frequency $\num = \Omegam/2\pi$ and displacement profile $\bv{Q}(\bv{r})$; each optical mode has a frequency $\nuo=\omegao/2\pi$ and electric field profile $\bv{E}(\bv{r})$.  Just as the optical mode volume, $\Vo = \int \mathrm{d}V \left(\frac{ \sqrt{\epsilon}\modulus{\bv{E}}}{\maxvec{\sqrt{\epsilon}E}}\right)^2$, describes the electromagnetic localization of the optical mode, the mechanical mode volume, $\Vm \equiv \rho \int \mathrm{d}V \left(\frac{\modulus{\bv{Q}}}{\maxvec{Q}}\right)^2$ (see App. \ref{appD}), describes the strain energy-averaged localization of the mechanical mode.  For both the localized optical and mechanical modes of the patterned beam cavity, the effective mode volume is less than a cubic wavelength.  The effective motional mass, being proportional to the mode volume ($\meff \equiv \rho \Vm$), is between 50 and 1000 femtograms for the mechanical modes shown in Fig.~\ref{fig:bands}c ($\rho_{Si}=2.33$ g/c$\textrm{m}^3$).    

\begin{figure*}[t]
\begin{center}
\includegraphics[width=0.85\columnwidth]{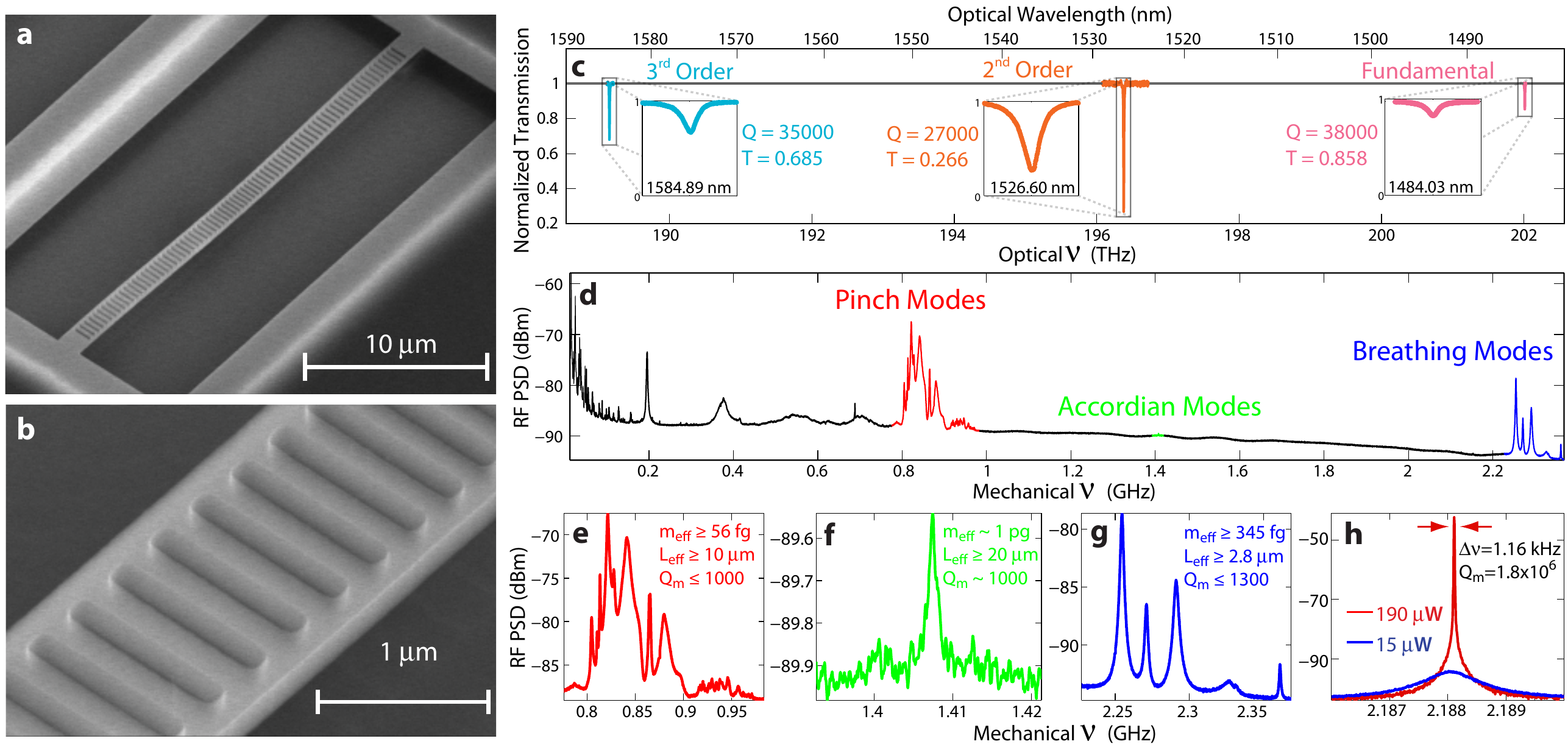}
\caption{\textbf{Photonic and phoninic crystal mode spectroscopy}. \textbf{a}, and \textbf{b}, show SEM images of the fabriced silicon nanobeam optomechanical crystal.  \textbf{c}, Optical spectroscopy of Device 1 with the taper waveguide in contact.  \textbf{d}, Mechanical spectroscopy of Device 1 with taper waveguide in contact.  \textbf{e}-\textbf{g}, Zoomed RF mechanical spectra of Device 1 showing pinch (red), accordian (green), and breathing (blue) modes.  \textbf{h}, Fundamental breathing mode RF spectra (of a third device) at low ($15$ $\mu$W) and high ($190$ $\mu$w) coupled optical power .  At high optical power the breathing mode is optically amplified through dynamical back-action of the radiation pressure force, with an above threshold effective $Q$-factor greater than $10^6$.} \label{fig:sems}
\end{center}
\end{figure*}

Drawing on recent work in the field of cavity optomechanics\cite{ref:TJK_KJV_Science,ref:Favero09}, we describe the coupling between optical and mechanical degrees of freedom (to lowest order) by an effective coupling length $\Leff \equiv (\frac{1}{\nuo}\deriv{\nuo}{\alpha})^{-1}$ (see App. \ref{appD}), where $\delta \nuo$ is the change in the frequency of an optical resonance caused by the mechanical displacement parameterized by $\alpha$.  For this work, $\alpha$ is defined as the maximum displacement that occurs \emph{anywhere} for the mechanical mode.  By definition then, the smaller $\Leff$, the larger the optical response for a given mechanical displacement.  $\Leff$ is also the length over which a photon's momentum is transferred into the mechanical mode as it propagates within the structure, and thus is inversely proportional to the force per-photon applied to the mechanical system.  

To calculate $\Leff$, we employ a perturbative theory of MaxwellÕs equations with respect to shifting material boundaries\cite{ref:Johnson_shifting_boundaries}.  The derivative $\deriv{\nuo}{\alpha}$ around some nominal position, where the optical fields are known, can be calculated \emph{exactly} without actually deforming the structure for a surface-normal displacement of the boundaries, $h(\alpha; \bv{r}) \equiv \bv{Q}(\bv{r}) \cdot \bv{\hat{n}} = \alpha \bv{q}(\bv{r}) \cdot \bv{\hat{n}}$, where $\bv{q}(\bv{r})=\bv{Q}(\bv{r})/\alpha = \mathrm{d}\bv{Q}(\bv{r})/\mathrm{d}\alpha$ is the unitless displacement profile of the mechanical mode, and $\alpha$ parameterizes the amplitude of the displacement.  Using this perturbative formulation of Maxwell's equations, we find

\begin{equation}\label{eq:Leff}
\frac{1}{\Leff} = \frac{1}{2}\frac{ \displaystyle \int \mathrm{d}A \left(\deriv{\bv{Q}}{\alpha} \cdot \bv{\hat{n}} \right) \left[\Delta \epsilon \modulus{\bv{E}_\parallel}^2 - \Delta (\epsilon^{-1}) \modulus{\bv{D}_\perp}^2 \right]}{\displaystyle \int \mathrm{d}V \epsilon \modulus{\bv{E}}^2}
\end{equation}

\noindent where $\bv{\hat{n}}$ is the unit normal vector on the surface of the unperturbed cavity, $\bv{D}(\bv{r}) = \epsilon(\bv{r})\bv{E}(\bv{r})$, $\Delta \epsilon = \epsilon_1 - \epsilon_2$, $\Delta (\epsilon^{-1}) = \epsilon_1^{-1} - \epsilon_2^{-1}$, $\epsilon_1$ is the dielectric constant of the periodic structure, and $\epsilon_2$ is the dielectric constants of the surrounding medium ($\epsilon_2 = \epsilon_0$ in this case).  This method of calculating the coupling provides a wealth of intuition about the nature of the coupling and can be used to engineer the structure for strong optomechanical coupling.

Figs.~\ref{fig:sems}a and ~\ref{fig:sems}b show scanning electron microscope (SEM) images of a fabricated silicon nanobeam with the parameters of Device 1 (see App. \ref{appA}).  The optical modes of the nanobeam are probed with a tapered and dimpled optical fiber\cite{ref:MichaelCP1} in the near-field of the defect cavity, simultaneously sourcing the cavity field and collecting the transmitted light in a single channel (see App. \ref{appB}).  Fig.~\ref{fig:sems}c shows the low-pass filtered optical transmission spectrum of Device 1 at low optical input power ($\sim 30$ $\mu$W).  The optical cavity resonances are identified by comparison to FEM modeling of the optical modes of the structure.  Looking in the radio frequency (RF) spectrum provides information about the mechanical modes of the structure, as mechanical motion gives rise to phase and amplitude modulation of the transmitted light.  Figs.~\ref{fig:sems}c-f show the measured photodetector RF power spectral density (PSD) of the optical transmission through the second order cavity resonance (this mode was used due to its deep on-resonance coupling).  A series of lower frequency modes can be seen in the spectra ($\sim 200$ MHz and harmonics), corresponding to compression of modes of the entire beam, followed by groups of localized phononic modes of the lattice at $850$ MHz (pinch), $1.41$ GHz (accordian), and $2.25$ GHz (breathing).  The transduced signal at low optical power corresponds to thermally-excited motion of the mechanical modes, and is inversely proportional to $\meff \Leff^2$ (see below).  At higher optical input power (see Fig.~\ref{fig:sems}h), optical excitation of regenerative mechanical oscillation\cite{ref:TJK_KJV_Science} of the breathing modes is possible due to the small mass and short optomechanical coupling length of the co-localized phonon and photon modes.     

\begin{figure}[t]
\begin{center}
\includegraphics[width=0.55\columnwidth]{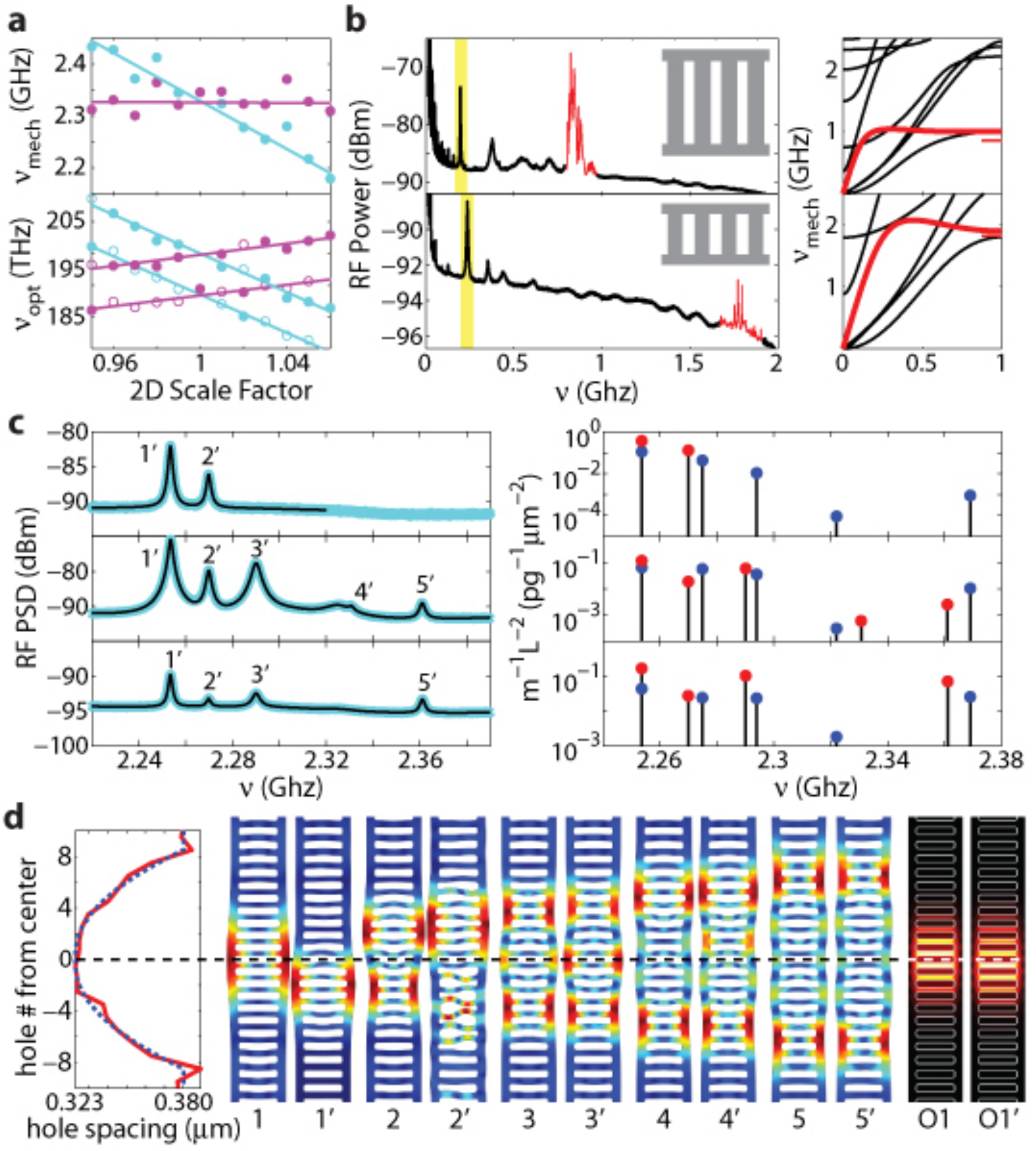}
\caption{\textbf{Phononic mode tuning and transduction}.  \textbf{a}, Geometric scaling (planar) of the fundamental breathing mode.  Device 1 is the device with scale factor 1.03.  The best linear least-squares fit lines in the top panel correspond to the mechanical frequency changing by $-0.9\%\pm0.2$\% per device; the normalized frequency changes by $-0.01\%\pm0.2\%$ per device.  The optical frequency of the mode used to make the mechanical measurement is filled (the other optical mode is open).  \textbf{b},  Engineering of pinch mode frequencies, showing two devices with pinch mode frequencies of 850 MHz and 1.75 GHz.  The mechanical band diagrams of each structure are shown to the right of the measured RF spectrum, with the pinch mode band highlighted in red. \textbf{c}, Transduction of breathing mode motion.   \textbf{d},  Ideal and actual (``primed") modes of the silicon nanobeam optomechanical crystal due to the ideal (dashed) and actual defect (solid, red).} \label{fig:spect}
\end{center}
\end{figure}

Fig.~\ref{fig:spect}a, top panel, shows how the frequency of the fundamental mechanical breathing mode scales with a uniform geometric scaling in the plane.  A series of 12 devices have been fabricated, identical except that the entire geometry in the plane is scaled incrementally by 1\% per device.  For each device, one of the first two optical modes is selected and used to measure the mechanical frequency of the fundamental breathing mode (cyan dots, top panel).  The frequencies plotted in magenta are the normalized frequencies, i.e. the bare frequencies (cyan) times the scale factor for the device.  Because the vibrations are entirely two-dimensional (in the plane), the frequency of the mechanical mode scales perfectly with the two-dimensional scale factor.  This is in contrast to the optical modes (Fig.~\ref{fig:spect}a, bottom panel), which clearly do not scale with the planar geometry, a result of the coupling of in-plane and vertical optical mode confinement (scaling in all three dimensions is thus required).  Since the planar scaling for the lattice-localized mechanical modes is trivial, this method could be used with a larger span of devices to measure the frequency dependence of the Young's (or bulk) modulus of the material.

Significant shifts in the frequency of the lattice-localized mechanical modes can be obtained through a non-uniform planar scaling.  Fig.~\ref{fig:spect}B shows the RF PSD for Device 1 and a second device, Decice 2, which has an essentially identical lattice constant, $\Lambda = 365$ nm, and total length, $L$, as compared to Device 1, but a considerably smaller width ($w = 864$ nm, $h_y = 575$ nm, $h_x = 183$ nm).  Simulations show that the pinch modes are the lowest-frequency group of localized and optomechanically-coupled mechanical modes in both structures (see right panels of Fig.~\ref{fig:spect}B).  Experimentally, the ratio of the localized pinch mode frequencies (highlighted in red) in these two devices is 1.749 GHz/805 MHz = 2.17.  The ratio of the frequency of the localized pinch-mode manifold, after accounting for the defect, is theoretically 1.826 GHz/846 MHz = 2.16.  It is interesting to note that the mechanical modes of the entire doubly-clamped beam (as opposed to the lattice-localized modes) depend very weakly on the structural differences between Device 1 and Device 2.  For instance, the second-order acoustic vibration mode of the nanobeam (highlighted in yellow in Fig.~\ref{fig:spect}B) has a frequnecy which should be $\frac{3 \pi}{2 L} \sqrt{\frac{E}{\left<\rho\right>}}$, where $E$ is Young's modulus and $\left<\rho\right>$ is the average linear density.  The frequency of this mode is measured to be 234 MHz/195 MHz = 1.20 times higher in Device 2 than for Device 1, which is in good agreement with the ratio $\sqrt{\langle \rho_1 \rangle/\langle \rho_2 \rangle} = 1.23$.  The difference between the change in the frequencies of the lattice-localized versus beam modes illustratres the independence of these two ``systems"; once the wavelength of the global beam modes approach the scale of the lattice periodicity, the vibrations become localized and behave independently of the global beam structure (such as the end clamps).

Fig.~\ref{fig:spect}c shows the RF optical transmission spectrum due to Brownian motion of the breathing modes of Device 1 (i.e., at low optical input power), for the three optical modes shown in Figs.~\ref{fig:bands}c and ~\ref{fig:sems}c.  Because the various optical modes have different spatial profiles, each mechanical mode has a \emph{different} $\Leff$ \emph{for each optical mode}.  The root-mean-square (rms) mechanical amplitude of a mode due to Brownian motion is $\left<\alpha^2\right>=k_B T/(\meff \Omega^2)$.  It can be shown analytically that the factor $1/(\meff\Leff^2)$ uniquely determines the transduction of the Brownian motion for these sideband-resolved optomechanical oscillations (see App. \ref{appE}).  To the right of each measured spectrum is the experimentally-extracted mechanical frequency and value of $1/(\meff \Leff^2)$, together with the values of these quantities obtained from the FEM model (using Eq. \ref{eq:Leff} to determine $\Leff$).  Good correspondence, in both frequency and transduced signal amplitude, is found across all optical and mechanical mode pairs.  In order to achieve this level of correspondence, imperfections in the fabricated structure are taken into account by extracting the geometry from high-resolution SEM images of the device and calculating the modified optical and mechanical modes (Fig.~\ref{fig:spect}d).  The resulting measured value for optomechanical coupling between the fundamental breathing and optical mode (assumming a FEM-calculated  motional mass of $\meff = 330$ fg) is $\Leff = 2.9$ $\mu$m, approaching the limit of the wavelength of light.  The sensitivity of the mechanical transduction of the fundamental breathing mode can be appreciated by comparing the mode's rms thermal amplitude at $T=300$~K, $\alpha_{th}=245$~fm, to its quantum zero-point motion of $\alpha_{zp}=3.2$~fm.  The sensitivity limit, as given by the background level in the middle panel of Fig.~\ref{fig:spect}c, is thus a factor of $\sim 7.5$ times that of the standard quantum limit.      

The loss of mechanical energy from confined mechanical modes of a phononic crystal can, in principal, be made arbitrarily low (and thus the mechanical $Q$ arbitrarily high) by including a large number of unit cells outside the localizing potential region.  Of course, other forms of mechanical damping, such as thermo-elastic damping, phonon-phonon scattering, or surface damping effects, would eventually become dominant\cite{ref:Roukes_intrinsic_dissipation}.  This makes optomechanical crystals ideal structures for studying these loss mechanisms.  The fundamental breathing mode of the 1D phononic crystal structure studied here, at $2.254$ GHz, has a room temperature mechanical $Q$ of $1300$ in air, and in contact with the taper waveguide (power-dependent measurements confirm that this mechanical Q is not enhanced by dynamical back-action).  This corresponds to a frequency-$Q$ product of $3 \times 10^{12}$ Hz, a value close to largest demonstrated to date\cite{ref:Weinstein_silicon_fq_record}.  Although further tests (as a function of temperature and lattice periods) are required to determine the contribution of various mechanical loss mechanisms, numerical simulations show that mode coupling between localized and leaky phonon modes exist in these 1D cavity structures and can significantly limit the $Q$-factor (see App. \ref{appC}).  This obstacle can be overcome in two-dimensional periodic slab structures, which have been shown to possess complete gaps for both optical and mechanical modes simultaneously\cite{ref:Thomas_simultaneous_gaps}.

The experimental demonstration of optomechanical coupling between 200 Terahertz photons and 2 Gigahertz phonons in a planar optomechanical crystal paves the way for new methods of probing, manipulating, and stimulating linear and non-linear mechanical and  optical interactions in a chip-scale platform.  As the study of quantum mesoscale mechanical oscillators has nearly become a reality\cite{ref:Arcizet_Nature_radiation_pressure_cooling_and_instability,ref:Gigan1,ref:Corbitt1,ref:ThompsonJD1,ref:TJK_close_to_ground_state}, high frequency mechanics will provide a distinct experimental advantage due to the lower thermal phonon occupancy.  In addition, optomechanical crystals with full phononic bandgaps provide a platform to decouple the direct decoherence (phonon leakage) of mechanical modes from their supports.  This could allow the preparation of mechanical vibrations with ultra-long lifetimes, the study of the intrinsic mechanical material losses, and narrow-linewidth Gigahertz frequency sources.  Optomechanical crystals could also be used as high-spatial resolution mass sensors; with $\meff = 62$~fg and $\num = 850$~MHz, the mass of a single Hemoglobin A protein ($\sim10^{-19}$~g) would change the frequency of the pinch mode by $700$~Hz, allowing sensitivity paralleling NEMS zeptogram mass sensors\cite{ref:Roukes_zeptogram}.

\begin{acknowledgments}
Funding for this work was provided by a DARPA seedling effort managed by Prof. Henryk Temkin, and through an EMT grant from the National Science Foundation.
\end{acknowledgments}

\appendix

\section{Fabrication}
\label{appA}

The optomechanical crystal nanobeam is formed in the 220 nm thick silicon device layer of a [100] Silicon-On-Insulator (SOI) wafer.  The pattern is defined in electron beam resist by electron beam lithography.  The resist pattern is transfered to the device layer by an inductively-coupled plasma reactive ion etch with a $\mathrm{C}_4\mathrm{F}_8$/$\mathrm{SF}_6$ gas chemistry.  The nanobeam is then undercut and released from the silica BOX layer by wet undercutting with hydrofluoric acid.

\section{Experimental Setup}
\label{appB}

\begin{figure}[htb]
\begin{center}
\includegraphics[width=0.7\columnwidth]{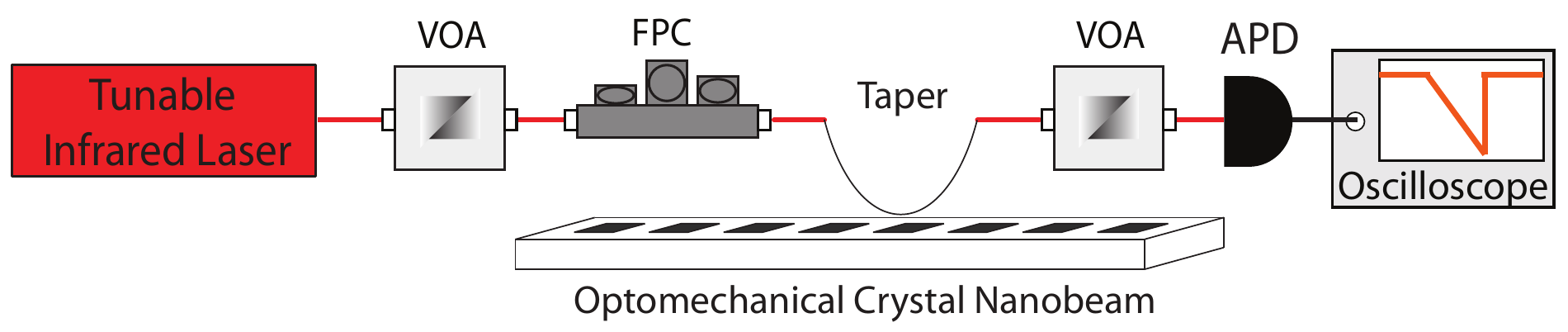}
\caption{Experimental setup used to measure optical, mechanical, and optomechanical properties of silicon optomechanical crystal nanobeam.} \label{fig:setup}
\end{center}
\end{figure}

The experimental setup used to measure the optical, mechanical, and optomechanical properties of the silicon optomechanical crystal nanobeam is shown in Fig.~\ref{fig:setup}.  The setup consists of a bank of fiber-coupled tunable infrared lasers spanning approximately 200 nm, centered around 1520 nm.  After a variable optical attenuator (VOA) and fiber polarization controller (FPC), light enters the tapered and dimpled optical fiber, the position of which can be controlled with nanometer-scale precision (although vibrations and static electric forces limit the minimum stable spacing between the fiber and device to about 50 nm).  The transmission from the fiber is (optionally) passed through another VOA and finally reaches an avalanche photodiode (APD) with a transimpedance gain of 11,000 and a bandwidth (3 dB rolloff point) of 1.2 GHz.  The APD has an internal bias tee, and the RF voltage is connected to the 50 Ohm input impedance of the oscilloscope.   The oscilloscope can perform a Fourier transform (FT) to yield the RF power spectral density (RF PSD).  The RF PSD is calibrated using a frequency generator that outputs a variable frequency sinusoid with known power.  

\section{Simulation Parameters}
\label{appC}

Modeling of both the optical and mechanical modes is done via finite element method (FEM), using COMSOL Multiphysics.  The following subsections provide the description of the method used to define the FEM model of the optomechanical crystal system.

\subsection{Extracting the geometry in the plane}

\begin{figure}[htb]
\begin{center}
\includegraphics[width=0.7\columnwidth]{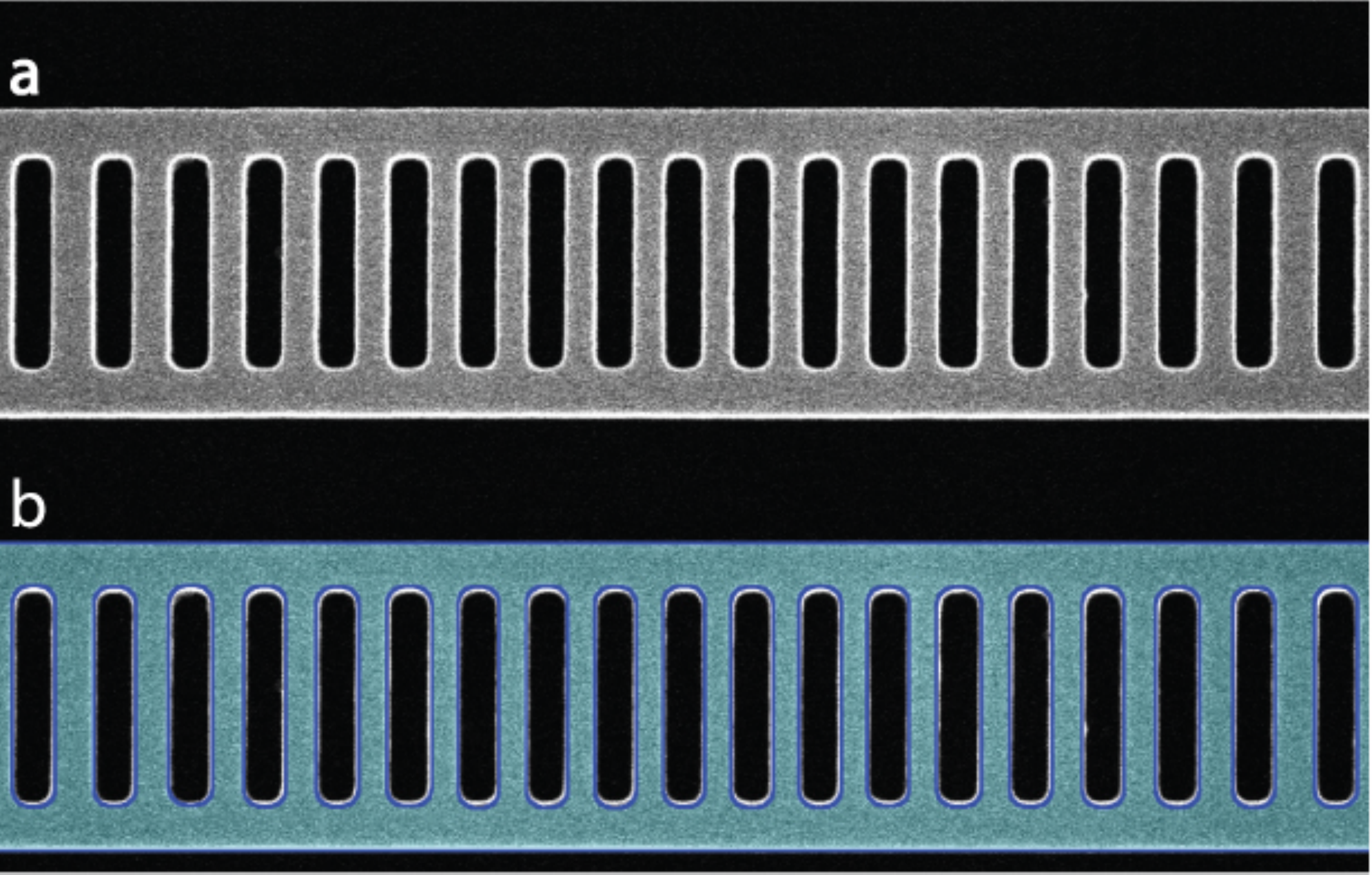}
\caption{(a) Scanning electron micrograph of fabricated silicon optomechanical crystal.  (b) Approximated geometry shown as blue overlay on SEM micrograph from (a).} \label{fig:overlay}
\end{center}
\end{figure}

To model the optomechanical crystal system, the geometry of the as-fabricated structure must be measured.   As the features are smaller than an optical wavelength, the measurements must be done by scanning electron microscope (SEM).  Figure~\ref{fig:overlay}a shows an ``eagle's-eye" high-resolution SEM micrograph of a portion of device 1, with the defect centered in the image.  

Digital line-scans of the micrograph are used to detect the edges of the geometry.  From the extracted edge positions, the geometry is approximated as a series of rectangular holes with two filleted ends inside of a rectangle (the beam), giving an approximate planar geometric representation of the structure shown as an overlay in Fig.~\ref{fig:overlay}b.  This geometric representation takes into account the size, position, and any curvature of each hole, giving an accurate approximation of the geometry.  In the defect region, each hole is given by its measured value.  Outisde the defect region, a series of holes is used to get the average hole shape, which is used in the model.

The SEM has been calibrated, and the dimesnions as measured by the SEM are too large by 5\%.  Thus, the entire planar geometry is uniformly scaled down by 5\%.  Since the lattice constant, $\Lambda$, is a center-to-center distance between features, it is not affected by erosion during processing, which makes it the most reliable measure of distance on the sample.  After applying the SEM calibration factor, the average lattice constant outside of the defect as measured by the SEM agrees with the value written by the electron beam lithography tool to better than 1\%.  Since the SEM and lithography tool are independent, this is yet another confirmation that the geometry has been measured correctly (the fine spectral features of the simulation are the other way to check the geometry measurements, after comparing to measured mechanical and optical spectra).

The SOI wafer thickness is specified as 220 nm by the manufacturer.  We will assume that the planar geometry extends uniformly into the vertical direction for the entire 220 nm, since the ``eagle's-eye" view used to measure the planar geometry does not capture any asymmetries in the vertical dimension.  These vertical asymmetries are much more difficult to extract without sacrificing the device (by focused ion beam, cleaving, etc.).

\subsection{Young's modulus and index of refraction}

The nanobeam structures are fabricated such that the long axis ($\hat{x}$) is parallel to the SOI wafer flat, which is oriented along $\left[110\right]$ ($\pm$0.5$^\circ$).  We   decompose the displacement field in FEM simulations along the crystal axes and find the majority of the strain energy is primarily stored in deformations along the family of equivalent directions specified by $\langle110\rangle$.  Because the strain for the modes of interest are primarily along $\langle110\rangle$, an isotropic elasticity tensor derived from a single Young's modulus and Poisson's ratio is appropriate for the current level of detail.  The index of refraction will also be treated as an isotropic scalar.

As the Young's modulus, $E$, and index of refraction, $n$, determine the phase velocity of the waves (and thus the frequency), they can be ``tuned" to make a single simulated frequency (mechanical for $E$ and optical for $n$) come out exactly as measured.  The free spectral range of the modes and the relative frequencies of different types of modes are determined by the details of the geometry, in conjunction with $E$ and $n$; so although a single frequency can always be made to match experiment exactly by scaling $E$ or $n$, the wider details of the spectrum are a more accurate reflection of whether the model is a good match to the experimental values.  

After accounting for the planar geometry and scale factor, the Young's modulus and index are tuned until the fundamental optical mode and the fundamental breathing mechanical mode each come out exactly as measured, which occurs for $E = 168.5$ GPa and $n = 3.493$.  These parameters, along with the measured geometry (as discussed above) yields a model that produces the values shown in Fig.~\ref{fig:spect}c.

\subsection{Optics: mode maps and modeling}

\begin{figure}[htb]
\begin{center}
\includegraphics[width=0.55\columnwidth]{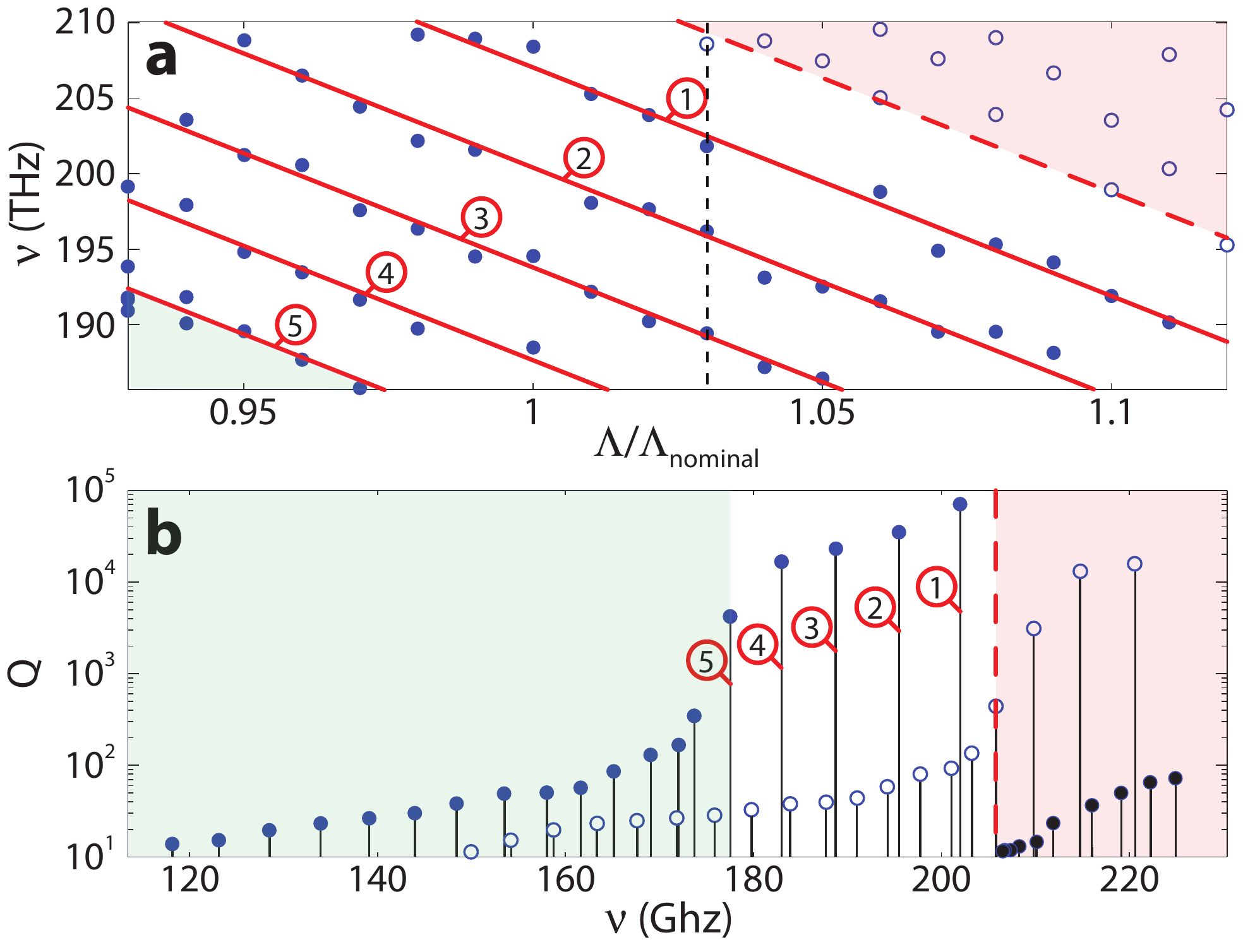}
\caption{(a)  Optical modes measured in a 200 nm laser wavelength span for a series of 20 devices.  (b)  Simulation of Device 1, which is the device with scaling factor 1.03 (dashed line).  Filled blue circles correspond to modes of the fundamental (valence) optical band; the region shaded blue corresponds to frequencies that are no longer within the defect potential barrier height (i.e. propagating modes).  Open blue circles correspond to transverse valence band modes.  Black circles correspond to conduction band modes.} \label{fig:optical_modes}
\end{center}
\end{figure}

Fig.~\ref{fig:optical_modes}a shows all the optical modes measured for a series of 20 devices, which are identical up to a uniform planar scaling that changes by 1\% per device (Device 1 is the device with scale factor 1.03).  Because of the limited laser range, only a limited number of modes can be measured on any given device.  By measuring this series of uniformly scaled devices, a large number of modes can be seen.  

The devices, taken together, display a number of conspicuous features.  First, the devices display a series of five modes with relatively high optical $Q$ (relative to the low-$Q$, waveguide-like modes that appear outside this range).  Second, the smallest devices show a number of low-$Q$ modes at frequencies below the fifth mode.  Finally, at high frequencies, the devices display another set of low-$Q$ modes, which are higher-$Q$ than waveguide-like modes but not as high-$Q$ as the other five modes.

These features are all consistent with the optical model of Device 1.  Figure~\ref{fig:optical_modes}b shows the simulated modes of Device 1, plotted as a function of their optical $Q$.  The simulation shows that the defect confines 5 modes, with a precipitous drop in $Q$ as the modal frequencies exit the defect potential (go above the energy barrier height).  The simulation also explains the series of modes higher in frequency, which are \emph{not} the conduction band modes.  These modes, indicated with open circles in both Fig.~\ref{fig:optical_modes}a and \ref{fig:optical_modes}b (as opposed to filled) circles, are the Hermite-Gauss ladder of modes with a single node transverse to the direction of propagation.  These modes have a lower effective index, which reduces their radiation-limited $Q$ relative to modes without transverse nodes.  Conduction band modes (which are not measured) are shown as filled black circles.

Even though the optical information provided by any single device would be difficult to unravel, the measurements of the series of devices coupled with simulation allow us to unambiguously identify the optical spectra of every device in the series.

\subsection{Modeling mechanical Q}\label{sec:Qmodel}

\begin{figure}[htb]
\begin{center}
\includegraphics[width=0.7\columnwidth]{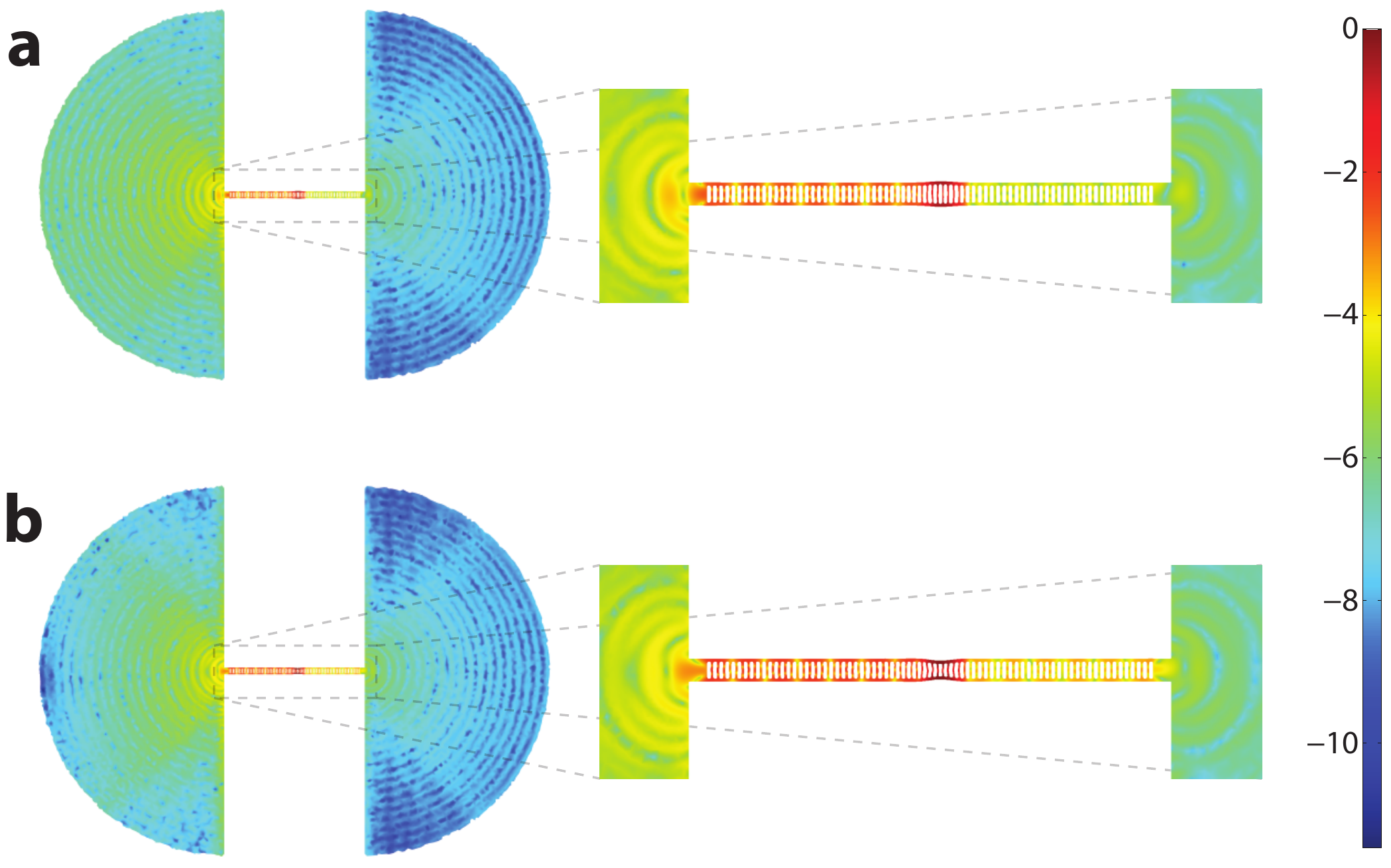}
\caption{(a) In-phase and (b) in-quadrature mechanical displacement field of fundamental breathing mode with absorbing ``pad".} \label{fig:mechq}
\end{center}
\end{figure}

Simulations of an isolated optomechanical crystal nanobeam (hard boundary conditions at the edges) show that the confined modes can couple to modes of longer wavelength, the effect of which is to produce a standing wave within the structure that is not attenuated outside the defect.  In the real structures, these long-wavelength modes will travel down the nanobeam and partially reflect at the contacts due to an effective impedance mismatch between the nanobeam and the bulk, with the rest of the power leaving the structure into the bulk.   Thus the isolated mode is coupled to a mode with long wavelength but identical frequency that can radiate into the surrounding ``bath".

To model the loss due to this resonant coupling to radiative modes, we include a large, semi-circular silicon ``pad" on each side of the nanobeam.  To make the pad act like a ``bath", we introduce a phenomenological imaginary part of the speed of sound in pad region; i.e., $\textrm{v}_{\textrm{pad}} \rightarrow \textrm{v}_{\textrm{Silicon}}(1+i \eta)$, where $\textrm{v} = \sqrt{E/\rho}$.  This creates an imaginary part of the frequency, and the mechanical $Q$ can be found by the relation, $Q_m = \Re{\num}/(2 \Imag{\num})$.  

By adding loss to the pad material, part of the power reflected at the contacts will be due to the change in the impedance from the absorption.  From this point of view, $\eta$ should be made as small as possible, since this contribution to the reflection coefficient is not present in the real system.  However, $\eta$ must also be large enough that the self-consistent solution includes a radiated wave that \emph{propagates} for a significant portion of the pad, which only happens if the wave is appreciably attenuated by the time it reflects from the edge of the simulation and returns to the contact.  Thus, the pad is made as large as possible, given computational constraints, and the absorption is increased until the $Q_m$ changes appreciably, which gives the threshold value for $\eta$ at which the reflectivity of the contacts has an appreciable contribution from the absorption.  The simulation is thus performed with a value of $\eta$ that produces a propagating wave in the pad without causing an artificial reflectivity at the contact; propagation in the pad is easily verified if the position of the nodes/antinodes swap between the in-phase and in-quadrature parts of the mechanical cycle (the nodes/antinodes of a standing wave are stationary).  Figs. \ref{fig:mechq}a and \ref{fig:mechq}b show the in-phase and in-quadrature (respectively) parts of the optical cycle, showing a propagating radiative mode in the pad.  The simulated values of $Q_m$ can be found in Table \ref{tabI}.  Most of the values of mechanical $Q$ calculated this way are in fair agreement with the measured values except for the fifth breathing mode, which, by this method, appears to have a $Q_m$ of over $22,000$; in reality, the $Q$ of this mode may be limited by coupling with a leaky mode caused by defects that are not modeled (such as vertical defects), material losses, or the presence of the taper waveguide.

Table \ref{tabI} summarizes the properties of the breathing mechanical modes.  Measured values are denoted with a tilde and extracted using techniques described below.  Simulated values are calculated using methods also described here in the appendices.
 
\begin{table*}[ht]
\caption{Measured and Simulated properties of the breathing mechanical modes.  Tildes indicate measured quantities.  The experimental effective lengths, $\Leff$, between each breathing mode and the first three  optical cavity modes are calculated using the experimentally extracted $\meff \Leff^2$ (see Fig.~\ref{fig:spect}c of main text) and dividing by the $\meff$ from the model.  See $\S$\ref{sec:Qmodel} for discussion on modeling $Q_m$.}
\label{tabI}
\begin{ruledtabular}
\begin{tabular}{llllllllllll}
Mode &$\num$ (GHz)&$\widetilde{\num}$ (GHz) & $\meff$ (fg) & ${}^1\Leff$ ($\mu$m) &$\widetilde{{}^1\Leff}$ ($\mu$m)&${}^2\Leff$ ($\mu$m) &$\widetilde{{}^2\Leff}$ ($\mu$m)&${}^3\Leff$ ($\mu$m) &$\widetilde{{}^3\Leff}$ ($\mu$m)&$Q_m$& $\widetilde{Q_m}$\\
\hline
1&  2.254  &2.254   &329   &4.9   &2.9   &6.4   &5.1   &7.8   &4.4   &1280   &2050 \\
2&  2.275  &2.270   &399   &7.1   &4.5   &6.2   &12    &9.5    &9.8   &1130   &1180 \\
3&  2.294  &2.290   &628   &11    &N/A   &6.2  &5.3   &7.7   &4.1   &613    &1290 \\
4&  2.322  &2.326   &704   &110  &N/A   &64   &49    &26    &N/A   &973    &387 \\
5&  2.369  &2.361   &665   &38    &N/A   &11   &25    &7.1   &4.7   &950    &21600 \\
\end{tabular}
\end{ruledtabular}   
\end{table*}

\section{Cavity Optomechanics: Definitions}
\label{appD}

For clarity and self-consistency, the following sections will present all the aspects of cavity optomechanics relevant to the experiments, in the same notation used in the main text.  Much of the information can be found in other places\cite{ref:TJK_KJV_review,ref:TJK_KJV_Science,ref:Karrai_nphoton_review,ref:Haus_book}.

\subsection{Relationship Between Fields and Their Corresponding Profiles and Amplitudes}

Cavity optomechanics involves the mutual coupling of an optical mode to a mechanical mode of a deformable structure.       

The optical mode is characterized by a resonant frequency $\omegao = 2 \pi \nuo$ and electric field $\bv{E}(\bv{r})$.  The mechanical mode is characterized by a resonant frequency $\Omegam = 2\pi \num$ and displacement field $\bv{Q}(\bv{r})$, where $\bv{Q}(\bv{r})$ is the vector displacement describing perpendicular displacements of the boundaries of volume elements.  The cavity optomechanical interactions of the distributed structure and its spatially-dependent vector fields can be reduced to a description of two \emph{scalar} mode amplitudes and their associated mode volumes, with the coupling of the amplitudes parameterized by a single coupling coefficient.  

The mode amplitude, $c$, and complex vector field profile, $\bv{e}(\bv{r})$, are defined such that the complex electric field is $\bv{E}(\bv{r})=c\bv{e}(\bv{r})$ (the physical field is given by the real part of $\bv{E}(\bv{r})e^{i\omega t}$).  For pedagogical reasons, the amplitude $c$ is normalized such that the time averaged electromagnetic energy is equal to $\modulus{c}^2$; i.e. $U = \modulus{c}^2 = \frac{1}{2}\int \mathrm{d}V \epsilon \modulus{\bv{E}}^2$.  This forces $\bv{e}$ to be normalized such that $1= \frac{1}{2}\int \mathrm{d}V \epsilon \modulus{\bv{e}}^2$.  In cavity quantum electrodynamics, one also defines an effective optical mode volume, $\Vo = \int \mathrm{d}V \left(\frac{ \sqrt{\epsilon}\modulus{\bv{E}}}{\maxvec{\sqrt{\epsilon} E}}\right)^2$, in order to gauge the strength of light-matter interactions.


The acoustic field amplitude, $\alpha$, and mode profile, $\bv{q}(\bv{r})$, are defined such that $\bv{Q}(\bv{r})=\alpha\bv{q}(\bv{r})$.  $\alpha$ is defined as the largest displacement that occurs anywhere for the mechanical field, $\bv{Q}(\bv{r})$, is $\alpha$.  It is important to note that this particular choice of $\alpha$ determines the mechanical mode effective volume and effective mass, $\Vm$ and $\meff \equiv \rho \Vm$, respectively.  In particular, this choice of $\alpha$ requires the complimentary definition: $\meff = \rho \int \mathrm{d}V \left(\frac{\modulus{\bv{Q}}}{\maxvec{Q}}\right)^2$.  To see this, note that the free evolution of the mechanical oscillator has, by definition, a time-independent total energy $E_{\textrm{mechanical}} = \frac{\meff}{2}(\Omega^2 \alpha^2 + \dot{\alpha}^2)$.  On the other hand, integrating the total energy of each volume element must \emph{also} give this same total energy.  If we pick the point in phase space at which all the mechanical energy is potential energy (i.e. the turn-around point), we must have that $E_{\textrm{mechanical}} = \frac{1}{2}\Omega^2 \int \rho |\bv{Q}(\bv{r})|^2 \mathrm{d}V = \frac{1}{2}\meff \Omega^2 \alpha^2$, or, in other words, $\meff \alpha^2 = \int \rho |\bv{Q}(\bv{r})|^2 \mathrm{d}V$.  One can arbitrarily choose the definition of the amplitude or the mass, but choosing one determines the other.  For a system like a confined mode of a phononic crystal defect cavity, where only a very small portion of the total mass undergoes appreciable motion, the most sensible choice of the mass is the amplitude-squared weighted density integral, which, as stated above, is the choice of mass associated with $\alpha = \mathrm{max}(|\bv{Q}(\bv{r})|)$.

\subsection{Coupling Between the Mechanical and Optical Modes}

The optomechanical coupling affects the optical mode by tuning its resonant frequency as a function of displacement, $\omegao(\alpha)$; whereas the coupling affects the mechanical mode by applying a force, which is expressed as a gradient of the cavity energy, $\mathrm{d}\modulus{c}^2/\mathrm{d}\alpha$.   The optical resonant frequency is usually expanded in orders of the (small) displacement, $\alpha$ around some equilibrium displacement, $\alpha_0$.

\begin{equation}
\omegao(\alpha) = \omegao\Big|_{\alpha=\alpha_0} + (\alpha-\alpha_0) \frac{\mathrm{d}\omegao}{\mathrm{d}\alpha}\Big|_{\alpha=\alpha_0} + (\alpha-\alpha_0)^2 \frac{\mathrm{d^2}\omegao}{\mathrm{d}\alpha^2}\Big|_{\alpha=\alpha_0}+...
\end{equation}

In the case that the terms higher than first order can be neglected, this equation simplifies to

\begin{equation}
\omegao(\alpha) = \omegao\Big|_{\alpha=\alpha_0} + (\alpha-\alpha_0) \frac{\mathrm{d}\omegao}{\mathrm{d}\alpha}\Big|_{\alpha=\alpha_0} \equiv \omegao + (\alpha-\alpha_0)g_{\mathrm{OM}} \equiv \omegao + (\alpha-\alpha_0)\frac{\omegao}{\Leff} \;,
\end{equation}

\noindent where $\omegao \equiv \omegao\Big|_{\alpha=\alpha_0}$ is the equilibrium resonance frequency of the optical mode, $g_{\mathrm{OM}} \equiv \frac{\mathrm{d}\omegao}{\mathrm{d}\alpha}\Big|_{\alpha=\alpha_0}$ is the derivative of the resonance frequency of the optical mode evaluated at equilibrium, and $\Leff^{-1} \equiv \frac{1}{\omegao}\frac{\mathrm{d}\omegao}{\mathrm{d}\alpha}\Big|_{\alpha=\alpha_0}$ is the \emph{effective length} of the optomechanical system, which is a universal parameter that relates the displacement relative to $\Leff$ to a relative change in optical frequency (i.e. $\alpha/\Leff = \delta \omegao/\omegao$).  It is simple to show that $\Leff$ is equal to the length of a Fabry-Perot cavity with one fixed mirror or the radius of a microtoroid or microdisk for the radial breathing mode.  For a zipper cavity or double-microdisk, $\Leff$ is a function of the spacing between the coupled elements that is exponential when the two elements are in the near-field of each other, with $\Leff$ approaching the wavelength of light as the spacing approaches zero.  

The optical force is the gradient of the optical mode energy

\begin{equation}\label{eq:Foptical}
\modulus{F_{\mathrm{optical}}} = \deriv{\modulus{c}^2}{\alpha} = \deriv{\modulus{c}^2}{\omegao} \deriv{\omegao}{\alpha} = \frac{\modulus{c}^2}{\omegao}\deriv{\omegao}{\alpha} = \frac{\modulus{c}^2}{\Leff}
\end{equation}

The perturbation theory of Maxwell's equations with shifting material boundaries\cite{ref:Johnson_shifting_boundaries} allows one to calculate the derivative of the shift of the resonant frequency of the optical modes of a structure in terms of some parameterization of a surface deformation \emph{perpendicular} to the surface of the structure.  Thus, if the result of a mechanical simulation is the displacement field, $\bv{Q}(\bv{r}) = \alpha \bv{q}(\bv{r}) \equiv \alpha \bv{Q}(\bv{r})/\maxvec{Q}$

\begin{equation}\label{eq:Leff}
\frac{1}{\Leff} = \frac{1}{2}\frac{ \displaystyle \int \mathrm{d}A \left(\deriv{\bv{q}}{\alpha} \cdot \bv{\hat{n}} \right) \left[\Delta \epsilon \modulus{\bv{E}_\parallel}^2 - \Delta (\epsilon^{-1}) \modulus{\bv{D}_\perp}^2 \right]}{\displaystyle \int \mathrm{d}V \epsilon \modulus{\bv{E}}^2}
\end{equation}

\noindent where $\bv{\hat{n}}$ is the unit normal vector on the surface of the unperturbed cavity, $\Delta \epsilon = \epsilon_1 - \epsilon_2$, $\Delta (\epsilon^{-1}) = \epsilon_1^{-1} - \epsilon_2^{-1}$, $\epsilon_1$ is the dielectric constant of the periodic structure, and $\epsilon_2$ is the dielectric constants of the surrounding medium.

\subsection{Coupled Equations of Cavity Optomechanics}

For an optical field $\bv{E}(\bv{r})$, which is described by the mode amplitude $c$, and a displacement (acoustic) field $\bv{Q}(\bv{r})$, described by mode amplitude $\alpha$, the optomechanical interaction of linear order (terms of order $\mathrm{d}^2 \omegao/\mathrm{d}\alpha^2$ and higher are neglected and $\alpha_0 \equiv 0$) of the fields is governed by the coupled differential equations 

\begin{eqnarray}\label{eq:diffeq_a}
&\dot{c}(t) = \left(-\frac{\kappa}{2}-i \omegao \left(1 \mp \frac{\alpha(t)}{\Leff}\right) \right)c(t) + i\sqrt{\frac{\kappa_e}{2}}s e^{-i \omega t} &\\ \label{eq:diffeq_alpha}
&\ddot{\alpha} + \Gamma_i\dot{\alpha} + \Omega^2 \alpha = \pm\frac{\displaystyle \left|c\right|^2}{ \displaystyle \meff \Leff} &
\end{eqnarray}

\noindent where $\kappa \equiv \kappai + \kappaext$, $\kappai \equiv \omegao/\Qopi$ is the intrinsic optical loss rate of the cavity; $\kappaext$ is the extrinsic coupling rate between the optical input field and the optical cavity field; $s$ is the amplitude of the input field, normalized such that $\modulus{s}^2 \equiv P_0$ is the optical input power impinging on the cavity; $\Omega$ is the acoustic resonance frequency of the cavity\footnote{Because the acoustic differential equation is second order, the damping coefficient that multiplies the velocity is the \emph{energy} loss rate, not the amplitude loss rate.  Thus $\Gamma$ is the linewidth of the power spectrum of the acoustic mode.  The distinction is important, as it can be confusing since it is $\kappa/2$ that appears as the damping coefficient in the first-order optical differential equation, but $\kappa$ is the linewidth (energy damping rate).}; $\Gamma \equiv \Omega/\Qmi$ is the intrinsic acoustic loss rate of the cavity; and $\meff \equiv \rho \Vac$ is the effective mass of the acoustic mode of the cavity, being the product of the mass density and the effective volume of the acoustic mode.  Here we use the convention that the optical resonance frequency is decreasing with increasing $\alpha$ and the optical force is the positive gradient of the cavity energy; this is the typical convention, but it is completely arbitrary\footnote{This is essentially a statement that the direction that positive $\alpha$ points is arbitrary.  The combination of the signs in the coupled differential equations, however, is not arbitrary.  One must change the sign of both to make physical sense, since, for instance, whether blue or red detuning induces mechanical gain will be changed if one is changed without the other.}.  

\section{Optical transduction of mechanical motion}
\label{appE}

One approximate method of solving the linear equations of optomechanics (\ref{eq:diffeq_a}) and (\ref{eq:diffeq_alpha}) is to treat the acoustic mode as a small perturbation of the optical mode, find the effect of the acoustic mode on the optical field, and then calculate what effect the perturbed optical mode has on the acoustic mode.  Starting with the premise that the displacement, $\alpha$, is sinusoidal; i.e.

\begin{equation}
\alpha(t) = \alpha_0 \sin(\Omega t) \;,
\end{equation}

\noindent the mode amplitude, $c$, is thus described by

\begin{equation}\label{eq:diffeq_sideband_method}
\dot{c}(t) = \left(-\frac{\kappa}{2}-i \omegao \left(1-\frac{\alpha(t)}{\Leff}\right) \right)c(t) + i\sqrt{\frac{\kappa_e}{2}}s e^{-i \omega t} \;.
\end{equation}

\subsection{Formal Solution}

The homogeneous solution to (\ref{eq:diffeq_sideband_method}) is

\begin{equation}
c_h(t) = C_0 \exp\left(\left( -\frac{\kappa}{2} - i \omegao \right)t  - i \frac{\alpha_0}{\Leff}\frac{\omegao}{\Omega}\cos(\Omega t)  \right) \;.
\end{equation}

The particular solution to (\ref{eq:diffeq_sideband_method}) can take the form $c_p(t) = C(t)c_h(t)$, satisfying

\begin{equation}
\dot{C}(t) = \frac{1}{c_h(t)} i\sqrt{\frac{\kappa_e}{2}}s e^{-i \omega t}  =i s\sqrt{\frac{\kappa_e}{2}} \exp\left(    \left(\frac{\kappa}{2} + i \omegao\right)t + i \frac{\alpha_0}{\Leff}\frac{\omegao}{\Omega}\cos(\Omega t)  -i \omega t \right) \;,
\end{equation}

\noindent where $C_0$ has been absorbed into $C(t)$.

Introducing the modulation index

\begin{equation}
\beta \equiv \frac{\alpha_0}{\Leff}\frac{\omegao}{\Omega} \;,
\end{equation}

\noindent the cosine part of the argument of the exponential can be expanded into Bessel functions as

\begin{equation}
\exp\left(  \pm i \beta \cos(\Omega t) \right) = \sum_{n=-\infty}^{+\infty} (\pm i)^n J_n(\beta)e^{i n \Omega t} \;,
\end{equation}

\noindent which is known as the Jacobi-Anger expansion.  This allows straight-forward integration of $C$; to wit,

\begin{equation}
C(t) = is\sqrt{\frac{\kappa_e}{2}}\sum_{n=-\infty}^{+\infty} \frac{i^n J_n(\beta)}{\frac{\kappa}{2} + i(n \Omega - \Delta)}e^{i \left(\frac{\kappa}{2} +n \Omega -\omega + \omegao \right)t} \;.
\end{equation}

\noindent Thus 

\begin{equation}\label{eq:cp}
c_p(t) = C(t)c_h(t) = is\sqrt{\frac{\kappa_e}{2}}\sum_{n=-\infty}^{+\infty} \frac{i^n J_n(\beta)}{\frac{\kappa}{2} + i(n \Omega - \Delta)}e^{i (n \Omega-\omega)t + i \beta \cos(\Omega t)} \;.
\end{equation}

\noindent The general solution is then $c(t) = c_h(t) + c_p(t)$.   As $c_h(t)$ is exponentially damped at rate $\kappa$, the general solution rapidly converges to $c_p(t)$, which is the steady-state solution.  This optical mode amplitude can thus be used to compute the various properties of the optomechanical system.

\subsection{The Transmission of an Oscillating Cavity}

The steady state power exiting the cavity is

\begin{equation}
s_{out} = i s e^{-i \omega t} - \sqrt{\frac{\kappa_e}{2}}c_p(t) \;,
\end{equation}

\noindent and thus

\begin{equation}
\modulus{s_{out}}^2 = \modulus{-is_{out}}^2 = \modulus{s e^{-i \omega t} + i \sqrt{\frac{\kappa_e}{2}}c_p(t)}^2 = \modulus{s}^2 + \frac{\kappa_e}{2}\modulus{c_p(t)}^2 - 2 \Imag{\sqrt{\frac{\kappa_e}{2}} c_p s^* e^{i \omega t} }
\end{equation}

\noindent with

\begin{equation}
- 2 \Imag{\sqrt{\frac{\kappa_e}{2}} c_p s^* e^{i \omega t} } = -\kappa_e\modulus{s}^2 \Re{\sum_{n,m}\frac{i^{(n-m)} J_n(\beta)J_m(\beta)}{\frac{\kappa}{2}+ i(n \Omega - \Delta)}e^{i (n+m)\Omega t}}
\end{equation}

\noindent and

\begin{equation}
\modulus{c_p(t)}^2 = \frac{\kappa_e}{2}\modulus{s}^2 \sum_{n,m}\frac{i^{(n-m)} J_n(\beta)J_m(\beta)}{\left(\frac{\kappa}{2}+ i(n \Omega - \Delta)\right)\left(\frac{\kappa}{2} - i(m \Omega - \Delta)\right)}e^{i (n-m)\Omega t} \;.
\end{equation}

\subsection{RF Spectrum of the First Order Sidebands}

The experimental arrangement in this work is such that the optical mode is observed by weakly populating the cavity with photons via the tapered optical fiber waveguide and then collecting the transmitted photons via the same waveguide.  The RF spectrum of the power transmitted, $\modulus{s_{out}}^2$, is obtained by simple photomixing on an APD.  This RF spectrum contains information about the mechanical modes because the mechanical modes modulate the optical mode via the optomechanical coupling.  The modulation manifests as a set of sidebands created in the cavity that are transmitted and detected at the APD.  If the amount of optical modulation, $\beta$, is small (i.e. $\beta \ll 1$), only the first sideband contributes, as $J_n(x \rightarrow 0) \approx \frac{1}{n!}\left(\frac{x}{2}\right)^n$.  In this case, only the product $J_0(\beta) J_{\pm1}(\beta)$ contributes to the signal, and we thus need only consider terms with $n=0,m=\pm1$ and $n=\pm1,m=0$ (terms with $n=0,m=0$ are DC and do not contribute to the RF spectrum).  Also note that, for $\beta \ll 1$, $J_0(\beta) \approx 1$, and $J_{\pm1}(\beta) \approx \pm \frac{\beta}{2}$.  As an example, for the breathing mode of a silicon nanobeam, one can have $\omegao/\Omega \approx 10^5$ and, with $\alpha$ given by the thermal amplitude of oscillation $\alpha = \sqrt{k_B T/\meff \Omega^2}$,  $\alpha/\Leff \approx 2\times 10^{-13}$ m$/5\times 10^{-6}$ m$ = 5\times 10^{-8}$, $\beta \approx 5\times 10^{-3}$.  Note that this condition is \emph{independent} of the degree of sideband resolution, which is $\Omega/(\kappa/2)$; the system can have small modulation in either the sideband-resolved or sideband-unresolved regimes.

In this small modulation approximation, the power oscillating at $\Omega$ is

\begin{eqnarray}
\frac{\modulus{s_{out,\Omega}}^2}{\modulus{s}^2 } &=& -\kappa_e \Re{ \frac{-i (\beta/2) e^{-i\Omega t}}{\frac{\kappa}{2} -i \Delta} + \frac{-i (\beta/2) e^{+i\Omega t}}{\frac{\kappa}{2} -i \Delta}  +   \frac{i (\beta/2) e^{+i\Omega t}}{\frac{\kappa}{2} + i (\Omega - \Delta)}    + \frac{i (\beta/2) e^{-i\Omega t}}{\frac{\kappa}{2} + i (-\Omega - \Delta)}     } \\
&&+ \left(\frac{\kappa_e}{2}\right)^2 \left\{     \frac{-i (\beta/2) e^{-i\Omega t}}{\left(\frac{\kappa}{2} - i  \Delta\right)\left(\frac{\kappa}{2} - i (\Omega - \Delta)\right)}    \right\}  \nonumber\\
&&+ \left(\frac{\kappa_e}{2}\right)^2 \left\{     \frac{-i (\beta/2) e^{+i\Omega t}}{\left(\frac{\kappa}{2} - i  \Delta\right)\left(\frac{\kappa}{2} + i (\Omega + \Delta)\right)}    \right\}  \nonumber\\
&&+ \left(\frac{\kappa_e}{2}\right)^2 \left\{     \frac{i (\beta/2) e^{+i\Omega t}}{\left(\frac{\kappa}{2} + i (\Omega - \Delta)\right)\left(\frac{\kappa}{2} + i \Delta\right)}    \right\}  \nonumber\\
&&+ \left(\frac{\kappa_e}{2}\right)^2 \left\{     \frac{i (\beta/2) e^{-i\Omega t}}{\left(\frac{\kappa}{2} - i (\Omega + \Delta)\right)\left(\frac{\kappa}{2} + i \Delta\right)}    \right\}  \nonumber\;.
\end{eqnarray}

This expression can be simplified by combining the terms proportional to $(\kappa_e/2)^2$ (those four terms are really just two terms plus their complex conjugates, which can be written as twice the real part of one term).  With this simplification, and collecting terms proportional to $\sin(\Omega t)$ and $\cos(\Omega t)$, one finds:

\begin{eqnarray} \label{eq:transmitted_power_separate_quadratures}
\frac{\modulus{s_{out,\Omega}}^2}{\modulus{s}^2 } = \cos(\Omega t)&& \Bigg[\frac{\kappa_e \beta}{2}  \left(  -\frac{2\Delta}{\left(\frac{\kappa}{2}\right)^2 + \Delta^2} - \frac{\Omega-\Delta}{\left(\frac{\kappa}{2}\right)^2 + (\Omega-\Delta)^2} + \frac{\Omega+\Delta}{\left(\frac{\kappa}{2}\right)^2 + (\Omega+\Delta)^2} \right) \\
&&+ \beta \frac{\kappa_e^2}{4} \left(   \frac{2 \kappa }{\kappa^2 + 4 \Delta^2} \right) \left( \frac{\Omega}{(\frac{\kappa}{2})^2+(\Omega-\Delta)^2} - \frac{\Omega}{(\frac{\kappa}{2})^2+(\Omega+\Delta)^2}   \right)  \Bigg] \nonumber \\
+\sin(\Omega t) && \Bigg[  \frac{\kappa_e \beta}{2} \left(   \frac{\kappa/2}{\left(\frac{\kappa}{2}\right)^2 + (\Omega-\Delta)^2}  -   \frac{\kappa/2}{\left(\frac{\kappa}{2}\right)^2 + (\Omega+\Delta)^2}      \right)   \nonumber \\ 
&&+ \beta \frac{\kappa_e^2}{4}	\left(   \frac{4 \Omega }{\kappa^2 + 4 \Delta^2} \right) \left( \frac{\Omega-\Delta}{(\frac{\kappa}{2})^2+(\Omega-\Delta)^2} - \frac{\Omega+\Delta}{(\frac{\kappa}{2})^2+(\Omega+\Delta)^2}   \right)   \Bigg] \nonumber \;.
\end{eqnarray}

\noindent If we say then, that

\begin{eqnarray}
\frac{\modulus{s_{out,\Omega}}^2}{\modulus{s}^2 } = \cos(\Omega t) A_{cos} + \sin(\Omega t) A_{sin} \;,
\end{eqnarray}

\noindent then the total electrical RF power detected at frequency $\Omega$ by a spectrum analyzer (i.e., in both quadratures) is proportional to

\begin{equation} \label{eq:transmitted_power_quadrature_sum}
\left(\modulus{s_{out,\Omega}}^2\right)^2 = \left(\modulus{s}^2\right)^2 \left(A_{cos}^2+A_{sin}^2 \right).
\end{equation} 

\subsection{Optical Forces}

Referring to equations~(\ref{eq:Foptical})~and~(\ref{eq:cp}), the optical force is equal to 

\begin{equation}
\modulus{c_p(t)}^2/\Leff = \frac{\kappa_e}{2 \Leff}\modulus{s}^2 \sum_{n,m}\frac{i^{(n-m)} J_n(\beta)J_m(\beta)}{\left(\frac{\kappa}{2}+ i(n \Omega - \Delta)\right)\left(\frac{\kappa}{2} - i(m \Omega - \Delta)\right)}e^{i (n-m)\Omega t} \;.
\end{equation}

\noindent We will again work in the limit of small modulation depth ($\beta \ll 1$).  In this caase, the DC force is found by considering the term for which $m=n=0$ (the next DC term is $m=n=\pm1$, which is of order $\beta^2$), which is simply proportional to the unperturbed cavity energy,

\begin{equation}
F_{\mathrm{optical,DC}} = \frac{1}{\Leff}\frac{2(\kappa_e/\kappa^2)}{1+4\left(\frac{\Delta}{\kappa}\right)^2} P_0 \;.
\end{equation}

\noindent In the limit of small modulation, just as in the consideration of the RF transmission spectrum, there will be four terms at frequency $\Omega$,

\begin{eqnarray}
F_{\mathrm{optical,\Omega}} &\equiv & F_Q \cos(\Omega t) + F_I \sin(\Omega t) \nonumber = F_Q \frac{\dot{\alpha}}{\Omega \alpha_0} + F_I \frac{\alpha}{\alpha_0}\\
&=& \modulus{s}^2\frac{\kappa_e \beta}{2 \Leff} \Bigg[ \cos(\Omega t)\left(   \frac{2 \kappa }{\kappa^2 + 4 \Delta^2} \right) \left( \frac{\Omega}{(\frac{\kappa}{2})^2+(\Omega-\Delta)^2} - \frac{\Omega}{(\frac{\kappa}{2})^2+(\Omega+\Delta)^2}   \right) \\
&& +\sin(\Omega t) \left(   \frac{4 \Omega }{\kappa^2 + 4 \Delta^2} \right) \left( \frac{\Omega-\Delta}{(\frac{\kappa}{2})^2+(\Omega-\Delta)^2} - \frac{\Omega+\Delta}{(\frac{\kappa}{2})^2+(\Omega+\Delta)^2}   \right)   \Bigg] \nonumber \;,\\
\end{eqnarray}

\noindent where $F_Q$ and $F_I$ are the in-quadrature and in-phase components of the force, respectively.

We can now rewrite the differential equation for the acoustic mode, Eq.~(\ref{eq:diffeq_alpha}), to include the effects of the optical mode.

\begin{equation}
\ddot{\alpha} + \Gamma_i \dot{\alpha} + \Omega^2 \alpha =\frac{1}{\meff}\left( F_Q \frac{\dot{\alpha}}{\Omega \alpha_0} + F_I \frac{\alpha}{\alpha_0} + F_{\mathrm{DC}}\right)
\end{equation}

\noindent Thus, we can rewrite the acoustic mode amplitude's differential equation as

\begin{equation}
\ddot{\alpha} + (\Gamma_i+\Gamma) \dot{\alpha} + \left(\Omega^2 + \delta\Omega^2\right) \alpha =\frac{1}{\meff}F_{\mathrm{DC}} \;,
\end{equation}

\noindent with

\begin{eqnarray}
\Gamma &\equiv& -\frac{1}{\meff \Omega \alpha_0}F_Q \label{eq:Gamma_definition}\\
\delta(\Omega^2) \approx 2\Omega \delta\Omega &\equiv& - \frac{1}{\meff \alpha_0}F_I  \;
\end{eqnarray} 

\begin{eqnarray}
\Gamma  &=& -\frac{ \omegao }{\Omega  \Leff^2 \meff    } \left(   \frac{2 \kappa_e \modulus{s}^2 }{\kappa^2 + 4 \Delta^2} \right) \left( \frac{\kappa/2}{(\frac{\kappa}{2})^2+\left( \Omega-\Delta \right)^2} - \frac{\kappa/2}{(\frac{\kappa}{2})^2+\left(\Omega+\Delta\right)^2}   \right) \\
\delta\Omega &=& -\frac{ \omegao}{ 2 \Omega \Leff^2 \meff}  \left(   \frac{2  \kappa_e \modulus{s}^2}{\kappa^2 + 4 \Delta^2} \right) \left( \frac{\Omega-\Delta}{(\frac{\kappa}{2})^2+(\Omega-\Delta)^2} - \frac{\Omega+\Delta}{(\frac{\kappa}{2})^2+(\Omega+\Delta)^2}   \right)
\end{eqnarray}

\subsection{Power transfer and effective temperature}

The power transfer between the optical and mechanical mode is (see, for instance, \cite{ref:TJK_KJV_review})

\begin{equation}
\left<P\right> = \left<F_{\mathrm{optical}}\cdot \dot{\alpha}\right> = \left< \alpha_0 \Omegam \left( F_Q \cos(\Omega t)^2 + F_I \sin(\Omega t)\cos(\Omega t)\right) \right> = \frac{\alpha_0 \Omegam}{2}F_Q \;.
\end{equation}

\noindent With equation \ref{eq:Gamma_definition}, we can thus write $ \left<P\right> = - \frac{\alpha_0^2}{2}\meff \Omegam^2 \Gamma = -\left<\alpha^2\right>\meff \Omega^2 \Gamma$.  The free evolution of the average mechanical energy, $E_m$ obeys $\ddt \left<E_m\right> = -\Gamma_i\left<E_m\right> +k_BT_R\Gamma_i$, where $T_R$ is the reservoir temperature.  This gives the expected steady-state result that $\left<E_m\right>  = k_B T_R$.  In the presence of the optical field, the loss rate of the mechanical system is modified.  However, the optical field does not modify the reservoir temperature\cite{ref:TJK_KJV_review}, which means that the evolution of the mechanical energy changes to $\ddt \left<E_m\right> = -(\Gamma_i+\Gamma)\left<E_m\right> +k_BT_R\Gamma_i$.  Thus the steady-state result gives an effective temperature $T_{eff} = \frac{\Gamma_i}{\Gamma_i + \Gamma}T_R$.  The effective temperature changes the total power of a mechanical mode, whereas the change in linewidth from $\Gamma$ just narrows (or broadens) the mechanical resonance.

\subsection{Calculating the Power Spectral Density}

The thermal amplitude of oscillation is defined by

\begin{equation}
\frac{1}{2} \meff \Omega^2 \alpha_{thermal}^2 = \frac{1}{2}k_B T \;.
\end{equation} 

\noindent Thus we can define a thermal modulation index given by

\begin{equation}
\beta_{thermal} = \frac{\alpha_{thermal}}{\Leff}\frac{\omegao}{\Omega} \;.
\end{equation}

\noindent Using equations~(\ref{eq:transmitted_power_separate_quadratures})~-~(\ref{eq:transmitted_power_quadrature_sum}) and $\beta_{thermal}$, one can find the total transmitted power oscillating at frequency $\nu = \Omega/(2\pi)$, $\modulus{s_{out,\nu}}^2$.  Only a fraction of this output power reaches the detector, and we'll call this $P_{@det,\nu} = \mu \modulus{s_{out,\nu}}^2$

The transimpedance gain of the detector, $G_{TI,0}$, converts optical power to a voltage; the gain has a simple pole, however, at frequency $\nu_{det}$.  Thus the total gain of the detector at the frequency of the oscillator, $\nu$, is $G_{TI}(\nu) = G_{TI,0}/\left(1+(\nu/\nu_{det})^2\right)$.  This voltage is then fed to the input of a buffered channel amplifier with unity gain and a simple pole at frequency, $\nu_{scope}$ (however, in post-processing, the oscilloscope flattens its own response; so that power spectral densities do not contain the oscilloscope's pole); then the voltage across the load resistor (input impedance $Z$) at the output of the channel amplifier is used to compute a ``power", such that 

\begin{equation}
P_{RF,\nu} = \left(\frac{\mu \modulus{s_{out,\nu}}^2 G_{TI,0}(\nu)}{\left(1+(\nu/\nu_{det})^2\right)}\right)^2/Z \;.
\end{equation}

\noindent Because the spectrum of a mechanical resonator is distributed over all frequencies, we must calculate how much power is in a given frequency interval, $\mathrm{d}\nu \equiv \mathrm{d}\Omega/(2\pi)$.  If the spectrum is a Lorentzian, we must have that
\begin{eqnarray}
P_{RF,\nu} = \int_{-\infty}^{\infty} S^2(\nu) \mathrm{d}\nu' = \int_{-\infty}^{\infty} \frac{S^2(\nu)}{1+\left(\frac{\nu' - \nu}{\delta\nu/2}\right)^2} \mathrm{d}\nu' = \pi S^2(\nu) \delta \nu/2 = \frac{\pi \nu S^2(\nu)}{2 Q_m} \;.
\end{eqnarray}

\noindent Thus the power spectral density at the peak is
\begin{equation}\label{eq:integrated_power}
S^2(\nu) = \frac{2 Q_m P_{RF,\nu}}{\pi \nu} \;.
\end{equation}

\noindent Note that an oscilloscope typically displays $S^2(\nu)\left(RBW\right)$, where $RBW$ is the resolution bandwidth (the reciprocal of the time record length of the FFT).

\subsection{Extracting the product $\meff \Leff^2$ from experimental RF spectra}

The spectra of Fig.~\ref{fig:spect}c of the main text are fit using a sum of Lorentzian lineshapes plus a quadratic background.  From the fit parameters, Eq. \ref{eq:integrated_power} allows simple extraction of the total RF power in each mechanical mode.  Equations~(\ref{eq:transmitted_power_separate_quadratures})~-~(\ref{eq:transmitted_power_quadrature_sum}) can then be used to calculate the product $\meff \Leff^2$ for each mode, albeit indirectly (since it is not simple to invert the equations).   Given $\kappa_e$, $\kappa$, $\Delta$, $\num$, and $\nuo$, which are all experimentally-measured parameters (and $\Delta$ is directly set to be the $\Delta$ which gives the maximum transduction) the product $\meff \Leff^2$ uniquely determines the transduced power due to Brownian motion in each mode.  Thus, by varying $\meff \Leff^2$, one can find the value for the product that gives the correct transduced power.


\end{document}